\documentclass[aps,prb,nobibnotes,reprint,twocolumn,superscriptaddress]{revtex4-1}

\usepackage{amsmath, amssymb, amsfonts}
\usepackage{array, booktabs, tabularx}
\usepackage{xcolor, graphicx, tikz}
\usepackage[caption=false]{subfig}
\usepackage{hyperref}  
\usepackage{braket}
\usepackage{bm}
\usepackage[ruled]{algorithm2e}
\usepackage[normalem]{ulem} 

\hypersetup{colorlinks = True, allcolors = {blue}}

\SetAlCapSty{textnormal} 

\newcommand{\kBTS}[2]{ \ket{B^{#1}_{#2}} }
\newcommand{\kredBTS}[2]{ \ket{\psi^{#1}_{#2}} }

\newcommand{\FC}[1]{ c_{#1}^\dagger }
\newcommand{\FA}[1]{ c_{#1} }

\newcommand{\GC}[1]{ \Gamma_{#1}^\dagger }
\newcommand{\PC}[1]{ P_{#1}^\dagger }
\newcommand{\PA}[1]{ P_{#1} }
\newcommand{\PN}[1]{ N_{#1} }
\newcommand{\SPN}[1]{ \bar{N}_{#1} }
\newcommand{\StrPC}[1]{ P_{p_1}^\dagger \cdots P_{p_{#1}}^\dagger }

\newcommand{\SP}[1]{ S_{#1}^+ }
\newcommand{\SM}[1]{ S_{#1}^- }

\newcommand{\SZ}[1]{ S_{#1}^z }

\newcommand{\SumOp}[2]{ \sum_{1 \leq p_1 < \cdots < p_{#2} \leq #1} }

\newcommand{\BTP}[2]{B^{#1}_{#2}}
\newcommand{\BSet}[2]{\Phi^{#1}_{#2}}
\newcommand{\ComCom}[1]{\mathcal{O}(#1)}

\def\su2{$su(2)$}
\def\spinhalf{\text{spin-1/2}~}

\newcommand{\Eq}[1]{Eq.~({#1})}
\newcommand{\Fig}[1]{Figure~{#1}}
\newcommand{\Alg}[1]{Algorithm~{#1}}
\newcommand{\Sec}[1]{Section~{#1}}
\newcommand{\Reference}[1]{Ref.~{#1}} 
\newcommand{\Appx}[1]{Appendix~{#1}}

\newcommand{\st}[1]{\ifmmode\text{\sout{\ensuremath{#1}}}\else\sout{#1}\fi}

\begin{document}


\title{Correlated pair ansatz with a binary tree structure}

\author{Rishab Dutta}
\affiliation{Department of Chemistry, Rice University, Houston, TX, USA 77005}

\author{Fei Gao}
\affiliation{Department of Physics and Astronomy, Rice University, Houston, TX, USA 77005}

\author{Armin Khamoshi}
\affiliation{Department of Physics and Astronomy, Rice University, Houston, TX, USA 77005}

\author{Thomas M. Henderson}
\affiliation{Department of Chemistry, Rice University, Houston, TX, USA 77005}
\affiliation{Department of Physics and Astronomy, Rice University, Houston, TX, USA 77005}

\author{Gustavo E. Scuseria}
\affiliation{Department of Chemistry, Rice University, Houston, TX, USA 77005}
\affiliation{Department of Physics and Astronomy, Rice University, Houston, TX, USA 77005}


\begin{abstract}

We develop an efficient algorithm to implement the recently introduced binary tree state (BTS) ansatz on a classical computer.
BTS allows a simple approximation to permanents arising from the computationally intractable antisymmetric product of interacting geminals and respects size-consistency. 
We show how to compute BTS overlap and reduced density matrices efficiently.
We also explore two routes for developing correlated BTS approaches: Jastrow coupled cluster on BTS and linear combinations of BT states.
The resulting methods show great promise in benchmark applications to the reduced Bardeen--Cooper--Schrieffer Hamiltonian and the one-dimensional XXZ Heisenberg Hamiltonian.

\end{abstract}


\maketitle


\section{Introduction} \label{sec: intro}

Electronic structure methods based on a Slater determinant are known to be inadequate for strongly correlated systems, motivating the design of efficient reference wavefunctions tailored for strong correlation. 
An essential guide in this endeavor may be the concept of seniority, \cite{Racah1943,RingBook} which is defined as the number of unpaired electrons in a Slater determinant for a given molecular orbital basis.
Seniority is not a symmetry of a many-fermion Hamiltonian in general. Nevertheless, it has been shown that the seniority-zero sector---the manifold of Slater determinants with all electrons paired---contains a significant amount of electron correlation for strongly correlated systems with an even number of electrons. \cite{Bytautas2011,Henderson2014}
However, even when the electronic structure problem is restricted to the seniority-zero sector, solving the corresponding exact state, the doubly occupied configuration interaction (DOCI) \cite{Weinhold1967,Veillard1967} has combinatorial complexity. 
Pair coupled cluster doubles is a low polynomial cost method that approximates DOCI very well under weak correlation, \cite{Stein2014,Henderson2014} where DOCI is close to the correct seniority-zero wave function, but fails under strong correlation for attractive pairing \cite{HendersonQCC2014} and for the repulsive Hubbard model, \cite{WahlenStrothman2018} where neither pCCD nor DOCI match the right answer.

Perhaps the simplest wavefunction beyond a single Slater determinant for approximating DOCI in all correlation regimes is the antisymmetrized geminal power (AGP), \cite{Coleman1965,Bratoz1965} 
equivalent to the number-projected Bardeen--Cooper--Schrieffer (PBCS) state. \cite{RingBook} 
However, introducing additional correlators on top of an AGP reference is necessary to accurately describe certain strongly correlated systems. \cite{Neuscamman2013,Zen2014,Henderson2019,Henderson2020,GRAGP2020,KhamoshiCC2021,Khamoshi2021,LCAGP2021,Khamoshi2023}
Similar conclusions have been drawn from the spin-AGP ansatz, where correlated methods reasonably describe strongly correlated \spinhalf Heisenberg models. \cite{Liu2023}
Thus, finding an ansatz that approximates DOCI better than AGP without going beyond the computational complexity expected from a reference wavefunction is desirable.

Some of the present authors have recently introduced a quantum state preparation algorithm for AGP and a variationally more flexible state by modifying the AGP quantum circuit. \cite{ESPQC2023} 
We named this ansatz the binary tree state (BTS) for reasons to be discussed below and developed a state preparation algorithm for BTS on a quantum computer. \cite{ESPQC2023}
In this work, we show that BTS can be efficiently implemented on a classical computer, thus introducing a quantum-inspired classical algorithm. 
Our findings are particularly interesting since we will show below that BTS is closely related to the antisymmetrized product of interacting geminals (APIG). \cite{Silver1969} 
Computing APIG expectation values is known to be computationally intractable. \cite{Johnson2013}
We demonstrate that BTS provides an excellent starting point for approximating the ground states of the reduced BCS Hamiltonian and the one-dimensional XXZ Heisenberg Hamiltonian with a computational complexity that only scales with the fourth power of the system size.
We also explore post-BTS correlated methods in this work since BTS matrix elements can be readily evaluated.

After we provide the necessary theoretical background in \Sec{\ref{sec: background}}, we describe BTS in detail in \Sec{\ref{sec: bts_structure}}, and present numerical results in \Sec{\ref{sec: bts_results}}.
We also consider correlating a BTS utilizing a Jastrow-style coupled cluster ansatz and exploring linear combinations of BT states in \Sec{\ref{sec: correlation}}.
Notably, going beyond a single BTS leads to exceptionally accurate results for the model Hamiltonians discussed herein.


\section{Binary tree state} \label{sec: bts}

\subsection{Background} \label{sec: background}


\begin{figure}[t]
\centering
\includegraphics[width=0.9\columnwidth]{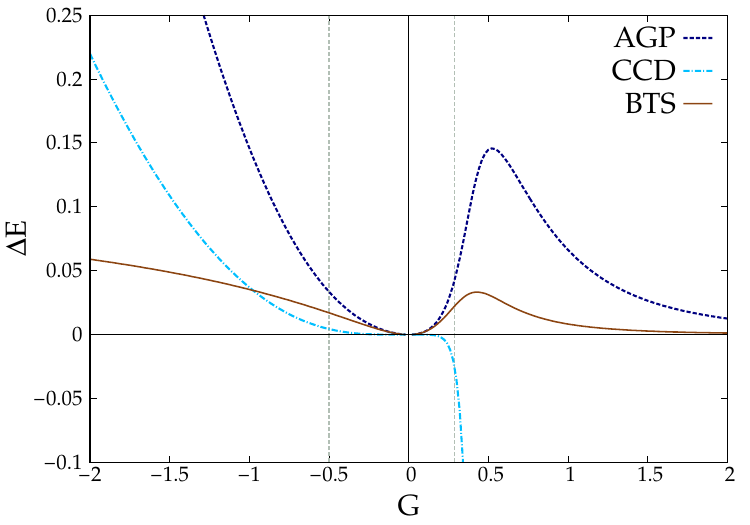}
\caption{
   Energy errors ($E_{\text{method}} - E_{\text{exact}}$) for the pairing Hamiltonian with $m = 16$ and $n = 8$, which has critical $G$ values $G_c = - 0.5$ for the repulsive regime and $G_c \sim 0.2866$ for the attractive regime.
   The horizontal axis plots the two-body interaction parameter $G$, with its critical $G$ values represented by the two vertical grey dotted lines. 
}
\label{fig: en_agp_bts_rbcs}
\end{figure}


The seniority-zero sector of the fermionic pairing channel can be described with pair-creation operators 
\begin{equation} \label{eq: pair_creation_op}
\PC{p} 
= \FC{p} \FC{\bar{p}}, 
\end{equation}
and their annihilation operator adjoints $\{ \PA{p} \}$, where $\{ p, \bar{p} \}$ represents paired orbitals in a given pairing scheme. \cite{Hurley1953}
The number operator 
\begin{equation} \label{eq: pair_num_op}
\PN{p} 
= \FC{p} \FA{p} 
+ \FC{\bar{p}} \FA{\bar{p}} 
\end{equation} 
and the pair-creation and annihilation operators satisfy \su2 commutation relations: 
\begin{subequations} \label{eq: pair_su2}
\begin{align}
[ \PA{p}, \PC{q} ]
&= \delta_{pq} \: ( I - \PN{q} ),
\\
[ \PN{p}, \PC{q} ] 
&= 2 \: \delta_{pq} \: \PC{q}.
\end{align}
\end{subequations}
In fermionic systems, there is another seniority-zero sector in the spin channel, which can be obtained by mapping pairing to spin ladder and $\SZ{}$ operators by conjugation of $\FC{\bar{p}} \FA{\bar{p}}$, defined in \Eq{\ref{eq: pair_creation_op}} and \Eq{\ref{eq: pair_num_op}} \cite{Anderson1958} 
\begin{subequations} \label{eq: pair_spin_mapping}
\begin{align}
\PC{p} 
&\leftrightarrow \SP{p}, 
\\
\PA{p} 
&\leftrightarrow \SM{p},
\\
\PN{p} 
&\leftrightarrow 2 \: \SZ{p} + I,
\end{align}    
\end{subequations}
for obtaining the \spinhalf \su2 algebra. This mapping allows direct application of the tools developed with pair operators to study \spinhalf systems. \cite{Liu2023}

The exact wave function in the seniority-zero Hilbert space is DOCI, and can be written with pairing operators as 
\begin{equation} \label{eq: doci}
\ket{\text{DOCI}}
= \SumOp{m}{n} C_{p_1 \cdots p_n} \: \StrPC{n} \ket{-},
\end{equation}
where $m$ is the number of paired levels, $n \: ( \leq m )$ is the number of pairs, and $\ket{-}$ is the physical vacuum. 
Equivalent expressions for spin systems can be obtained using the mapping defined in \Eq{\ref{eq: pair_spin_mapping}}. \cite{Liu2023}
Solving for the DOCI coefficients $\{ C_{p_1 \cdots p_n} \}$ scales combinatorially with system size.
Our aim is the practical design of computationally tractable wavefunctions that reasonably approximate DOCI.
One particularly appealing and conceptually simple approach towards DOCI is APIG
\begin{equation} \label{eq: apig}
\ket{\text{APIG}}
= \prod_{\mu = 1}^n \: \GC{\mu} \ket{-}, 
\end{equation}
where the operators $\{ \GC{\mu} \}$ in \Eq{\ref{eq: apig}} create geminals \cite{Surjan1999,Tecmer2022} in their natural orbital basis \cite{Lowdin1955,Henderson2020}
\begin{equation} \label{eq: geminal}
\GC{\mu}
= \sum_{p = 1}^m \: \eta_p^\mu \: \PC{p},
\end{equation}
where the scalars $\{ \eta_p^\mu \}$ are the geminal coefficients. 
We have assumed the geminal coefficients to be real-valued in this work; however, they can be complex-valued in general. \cite{Liu2023}
Expanding the APIG wave function reveals that the coefficients in the DOCI expansion are combinatorial functions known as permanents 
\begin{equation} \label{eq: apig_permanent}
C_{p_1 \cdots p_n}^{\text{APIG}}
= \sum_{\sigma \in \mathsf{S}_n} \eta_{\sigma_{p_1}}^1 \cdots \eta_{\sigma_{p_n}}^n,
\end{equation}
where $\mathsf{S}_n$ is the symmetric group over the symbols $\{ p_1, \cdots, p_n \}$ and $\sigma$ are the corresponding permutations.
Computing the permanent of a matrix is generally intractable, \cite{Valiant1979,Glynn2010} thereby making a general APIG an impractical method. 


\begin{figure*}[t]
\centering
\includegraphics[width=0.9\columnwidth]{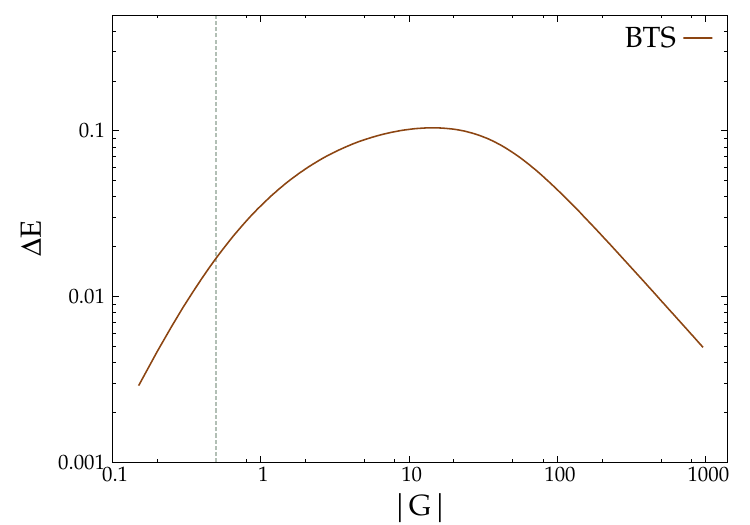}
\hskip3ex
\includegraphics[width=0.9\columnwidth]{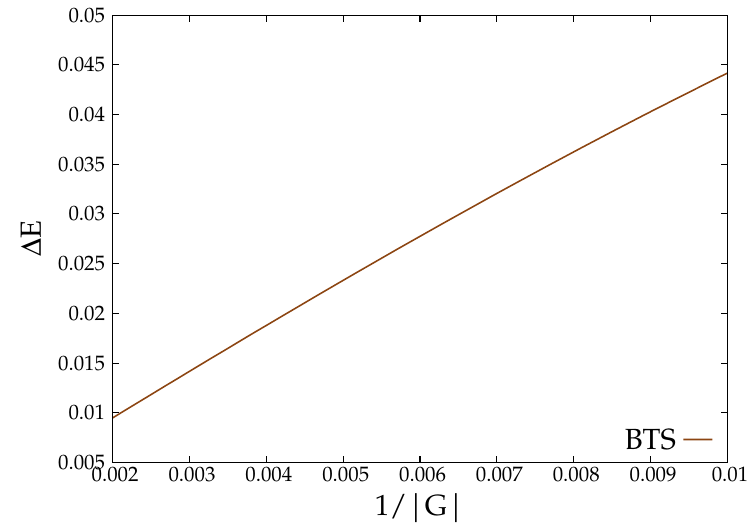}
\caption{
   Energy accuracy of BTS for the pairing Hamiltonian in the repulsive regime.
   The system has $m = 16$ and $n = 8$ with critical $G$ value at $G_c = - 0.5$. 
   Left panel: Energy error of BTS ($E_{\text{BTS}} - E_{\text{exact}}$) is plotted versus $| G |$ values. 
   A doubly logarithmic scale is chosen for the axes and the vertical grey line represents the $G_c$ value.
   Right panel: Energy error of BTS is plotted for small $1 / | G |$ values.
}
\label{fig: en_bts_rbcs_repul_m16n8}
\end{figure*}


Practical approximations of APIG often put constraints on the $\bm{\eta}$ geminal matrix structure and have been compared in \Reference{\citenum{Limacher2013}}. 
The antisymmetrized product of rank-$2$ geminals (APr2G) assumes the $\{ \eta_p^j \}$ elements have a special functional form such that the resulting permanents in \Eq{\ref{eq: apig_permanent}} are a ratio of determinants. \cite{Johnson2013,Limacher2013,Johnson2020,Fecteau2022,Moisset2022} 
The antisymmetrized product of strongly-orthogonal geminals (APSG), \cite{Hurley1953,Surjan2012} or its special case, generalized valence bond (GVB), \cite{Goddard1973} design a set of orthogonal geminals by expanding them in disjointed subspaces of molecular orbital basis, \cite{Arai1960,Lowdin1961} leading to a block structure of the $\bm{\eta}$ matrix.
A particularly simple approximation to APIG is AGP, where all the geminals are the same 
\begin{equation} \label{eq: agp}
\ket{\text{AGP}}
= \frac{1}{n!} \: \big( \GC{} \big)^n \ket{-}.
\end{equation}
As a result, the geminal index, $\mu$, is dropped from the $\GC{\mu}$ operator of \Eq{\ref{eq: geminal}} and the APIG geminal matrix $\bm{\eta}$ of \Eq{\ref{eq: apig}} reduces to a vector. 
Thus, the AGP expansion 
\begin{equation} \label{eq: agp_as_doci}
\ket{\text{AGP}}
= \SumOp{m}{n} \eta_{p_1} \cdots \eta_{p_n} \: \StrPC{n} \ket{-}   
\end{equation}
leads to a monomial approximation of the DOCI coefficients. \cite{Sangfelt1981,Khamoshi2019} 

We now take a subtly different route inspired by the AGP in approximating APIG.
We consider the BTS ansatz  
\begin{equation} \label{eq: bts}
\kBTS{m}{n}
= \SumOp{m}{n} \eta_{p_1}^1 \cdots \eta_{p_n}^n \: \StrPC{n} \ket{-},
\end{equation}
which maintains the monomial approximation of DOCI coefficients like AGP of \Eq{\ref{eq: agp_as_doci}} while retaining more variational flexibility. \cite{ESPQC2023}
Indeed, AGP is a special case of BTS, and thus, BTS is variationally bounded from above by AGP, i.e., $ E_{\text{AGP}} \geq E_{\text{BTS}} \geq E_{\text{DOCI}} $.
The structure of the BTS coefficients, as discussed in \Appx{\ref{app: btp}}, ensures that each APIG permanent expansion of \Eq{\ref{eq: apig_permanent}} becomes a monomial for BTS.
It should be noted that BTS can be thought of as a wavefunction resulting from the monomial approximation of DOCI coefficients, without referring to its relation with APIG.
Indeed, BTS can not simply be understood as a special case of APIG, and we have elaborated this point in \Appx{\ref{app: apig_and_bts}} with a specific example. 

\subsection{Structure of BTS} \label{sec: bts_structure}


\begin{figure*}[t]
\centering
\includegraphics[width=0.9\columnwidth]{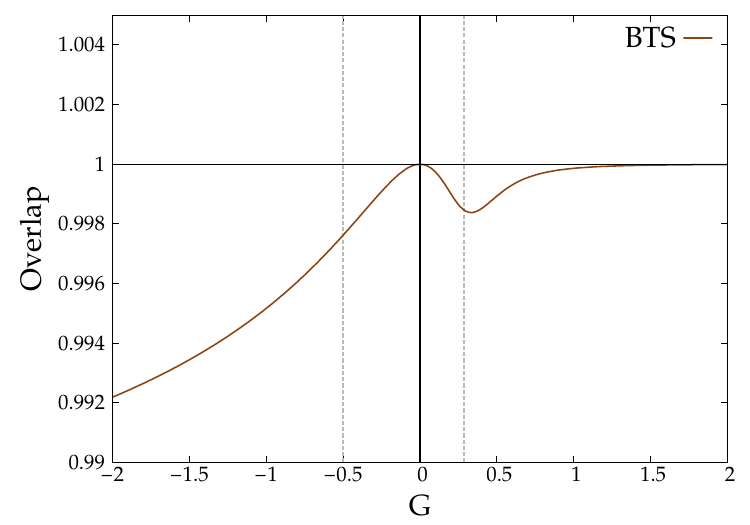}
\hskip3ex
\includegraphics[width=0.9\columnwidth]{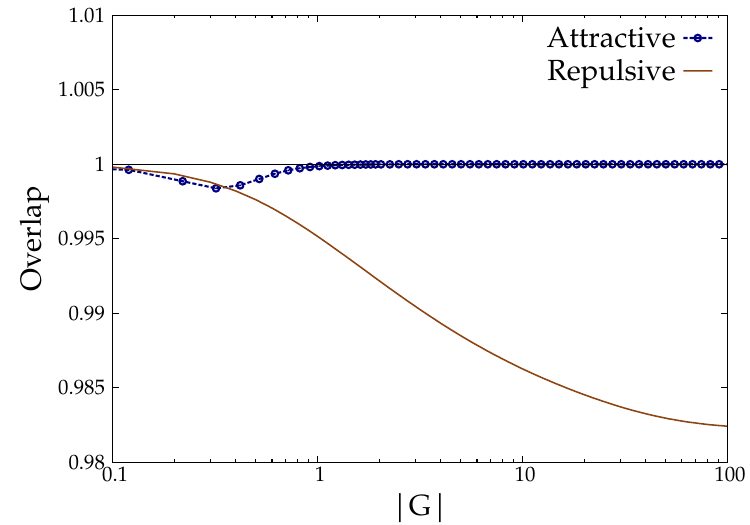}
\caption{
   Overlap of BTS with the exact wavefunction, $\braket{DOCI | BTS}$, for the pairing Hamiltonian with the two-body interaction parameter $G$. 
   The system has $m = 16$ and $n = 8$ with critical $G$ values $G_c = - 0.5$ for the repulsive regime and $G_c \sim 0.2866$ for the attractive regime.
   Left panel: Overlap values for a moderately large range of $|G|$ values.  
   The critical $G$ values are represented by the two vertical grey dotted lines. 
   Right panel: Overlap values for extremely large $|G|$ values.
   A logarithmic scale is chosen for the horizontal axis. 
}
\label{fig: ov_agp_bts_rbcs_m16n8}
\end{figure*}


\begin{figure*}[t]
\centering
\includegraphics[width=0.9\columnwidth]{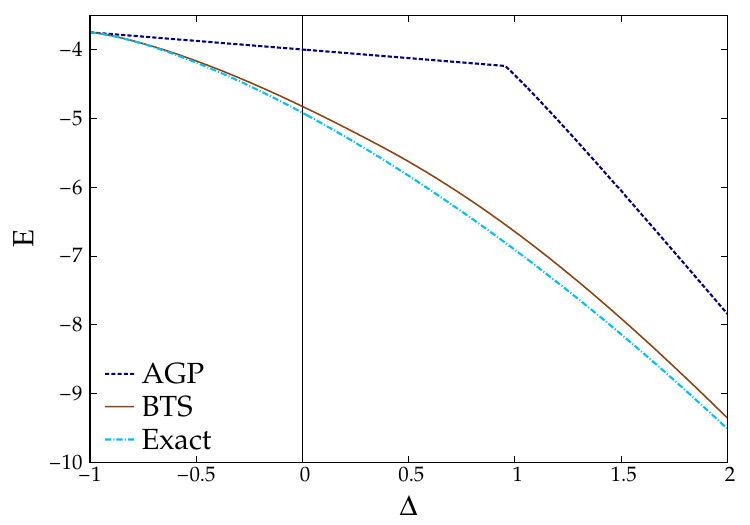}
\hskip3ex
\includegraphics[width=0.9\columnwidth]{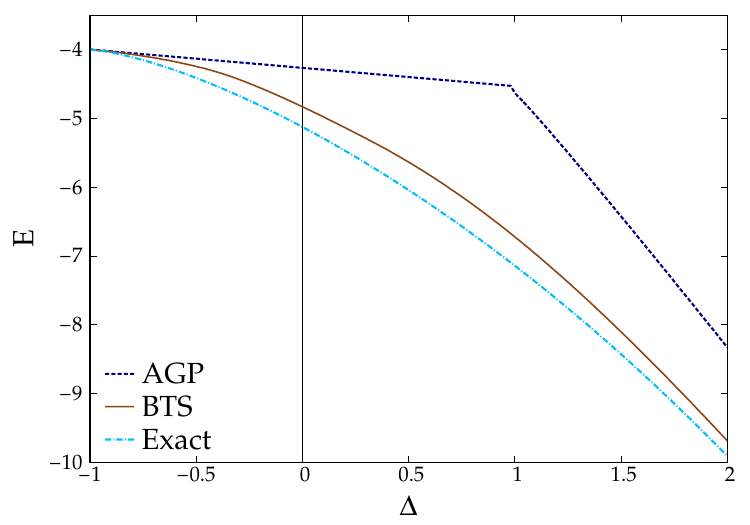}
\caption{
    Ground state energies for the one-dimensional XXZ Hamiltonian with 16 sites, $S_z = 0$.
    The horizontal axis represents the anisotropy parameter $\Delta$. 
    The left and right panels represent energies for open and periodic boundary conditions, respectively.
}
\label{fig: en_xxz_m16n8}
\end{figure*}


The BTS of \Eq{\ref{eq: bts}} has a particular polynomial structure, which we named the binary tree polynomial (BTP) in \Reference{\citenum{ESPQC2023}}.
A BTP of order $m$ and degree $n$ is a combinatorial function of an $n \times m$ matrix \textbf{A}:
\begin{equation} \label{eq: btp}
\BTP{m}{n} (\mathbf{A})
= \SumOp{m}{n} A_{p_1}^1 \cdots A_{p_n}^n,     
\end{equation}
where $\BTP{m}{n} = 0$ if $m < n$ and $\BTP{m}{0} = 1$ are also assumed.
The matrix elements $\{ A_p^\mu \}$ that do not appear in \Eq{\ref{eq: btp}} can be assumed to be zero.
Any BTP $\BTP{m}{n}$ can be computed in $\ComCom{m n}$ time, \cite{ESPQC2023}
which we discuss in more detail in \Appx{\ref{app: btp}}. 

The BTS of \Eq{\ref{eq: bts}} can be understood as a BTP of pair creation operators acting on the physical vacuum
\begin{equation} \label{eq: bts_as_btp}
\kBTS{m}{n}
= \BTP{m}{n} (\mathbf{A}) \ket{-}, 
\end{equation}
where we define $A_p^j = \eta_p^j \PC{p}$.
The BTS is normalizable, and its overlap is the following BTP \cite{ESPQC2023}
\begin{align} \label{eq: bts_overlap}
\braket{\BTP{m}{n} | \BTP{m}{n}}
= \SumOp{m}{n} | \eta_{p_1}^1 |^2 \cdots | \eta_{p_n}^n |^2, 
\end{align}
which allows its efficient computation. 
The following recursion relation exists for any BTS \cite{ESPQC2023} 
\begin{equation} \label{eq: bts_terminal_recursion}
\kBTS{p}{j} 
= \eta_p^j \PC{p} \kBTS{p - 1}{j - 1} 
+ \kBTS{p - 1}{j},  
\end{equation}
where $1 \leq j \leq n$ and $1 \leq p \leq m$ are defined based on their parent $\kBTS{m}{n}$ state.  
The binary tree structure of \Eq{\ref{eq: bts_terminal_recursion}} justifies the name of BTS. 
We formulate efficient algorithms for computing BTS reduced density matrices based on \Eq{\ref{eq: bts_terminal_recursion}} in \Appx{\ref{app: bts_rdms}}.

\subsection{Benchmarks} \label{sec: bts_results}

\begin{figure*}[t]
\centering
\includegraphics[width=0.9\columnwidth]{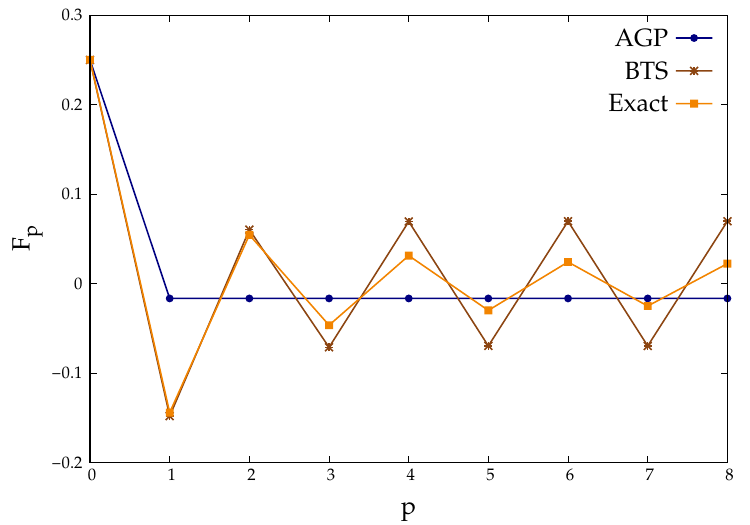}
\hskip3ex
\includegraphics[width=0.9\columnwidth]{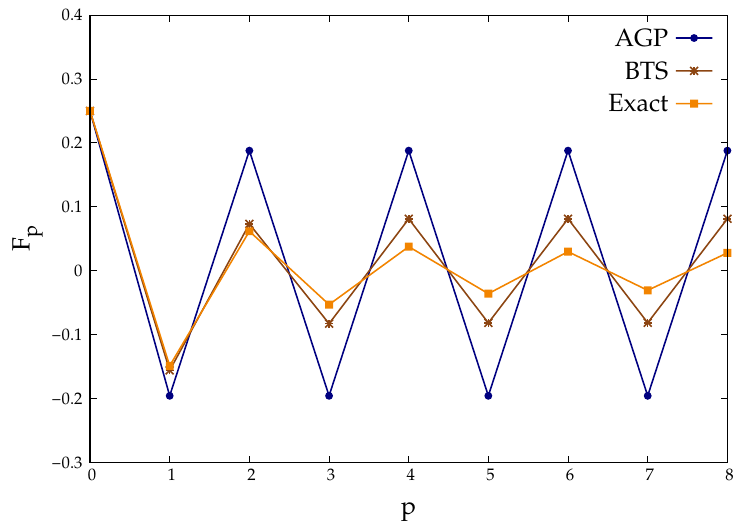}
\caption{
    Comparison of $S_z$-$S_z$ correlation functions for the one-dimensional XXZ Hamiltonian with 16 sites, $S_z = 0$, and periodic boundary conditions.
    The left and right panels are for $\Delta = 0.9$ and $\Delta = 1.0$, respectively.
    The vertical-axes plot the $S_z$-$S_z$ correlation function defined in \Eq{\ref{eq: sz_sz_correlation}} for varying site distances.
}
\label{fig: spin_xxz_pbc_m16n8}
\end{figure*}


We benchmark the BTS ansatz on two Hamiltonians for which DOCI is exact. 
The variational optimization of the BTS trial energy 
\begin{equation} \label{eq: en_bts}
E 
= \min_{\bm{\eta}} \frac{ \braket{\BTP{m}{n} | H |\BTP{m}{n}} }{ \braket{\BTP{m}{n} | \BTP{m}{n}} }  
\end{equation}
can be computed with $\ComCom{m^4}$ complexity following \Appx{\ref{app: btp_computation}} and \Appx{\ref{app: bts_rdms}}. 

The first Hamiltonian we study is the reduced Bardeen--Cooper--Schrieffer Hamiltonian \cite{BCS1957}  
\begin{equation} \label{eq: pairing_ham}
H 
= \sum_p^m \: \epsilon_p \: \PN{p} 
- G \: \sum_{pq}^m \: \PC{p} \PA{q}, 
\end{equation}
where $G$ is the two-body interaction parameter, and a constant spacing between $\{ \epsilon_p \}$ is assumed.
The Hamiltonian in \Eq{\ref{eq: pairing_ham}} belongs to a family of pairing Hamiltonians that can be exactly solved, \cite{Richardson1963,Richardson1964,Gaudin1976} 
and has been applied to study superconducting metallic grains. \cite{Dukelsky2004}
From now on, we will refer to \Eq{\ref{eq: pairing_ham}} as simply the pairing Hamiltonian.  
Electronic structure methods based on a single (number-conserving) Slater determinant fail to describe the attractive physics of the pairing Hamiltonian except at small $G$ values. \cite{HendersonQCC2014,Henderson2015,GRAGP2020} 
Indeed, the mean-field solutions break particle number and spin symmetry in the attractive and repulsive regimes of the pairing Hamiltonian, respectively. 
The $G$ values that correspond to these symmetry-breaking points are the so-called critical $G$ values.
AGP is known to be qualitatively accurate for the pairing Hamiltonian in the attractive interaction regime  \cite{Braun1998,Degroote2016,Dukelsky2016} and is exact in the large attractive $G$ limit of \Eq{\ref{eq: pairing_ham}}. \cite{Henderson2019}
However, AGP is less accurate in the moderately attractive and strongly repulsive regimes. 

\Fig{\ref{fig: en_agp_bts_rbcs}} shows that BTS energies are consistently more accurate than AGP. 
Specifically, BTS performs exceptionally well for the attractive regime of the pairing Hamiltonian, where traditional methods like coupled cluster doubles (CCD) based on the Hartree--Fock state are known to break down. 
For the moderately large repulsive regime, BTS has relatively more energy error, but still performs better than AGP and CCD. 
However, as shown in the left panel of \Fig{\ref{fig: en_bts_rbcs_repul_m16n8}}, the energy errors start to turn over so that BTS approaches the exact energy in the large negative $G$ limit of the pairing Hamiltonian.  

To investigate the asymptotic behavior further, we have computed the overlap of BTS and the exact ground state, as shown in \Fig{\ref{fig: ov_agp_bts_rbcs_m16n8}}. 
It turns out that the BTS does not reach the exact wavefunction at the limit $G \rightarrow -\infty$. 
This is due to the fact there is a high degree of degeneracy in the ground states of the pairing Hamiltonian at $G = -\infty$. \cite{Johnson2023}
We show in \Appx{\ref{app: CGtoBTS}} that BTS has a coefficient structure that can be mapped to one of the exact ground states at $G = - \infty$, while its behavior for $G \rightarrow - \infty$ can be explained by using arguments from perturbation theory.
All exact eigenfunctions of the pairing Hamiltonian are of APIG form (and in fact of APr2G form, with coefficients which permit easy computation of permanents), so the BTS overlap with the exact ground state in \Fig{\ref{fig: ov_agp_bts_rbcs_m16n8}} also measures how well BTS approximates a constrained APIG for this problem. 

The second model for our benchmark is the one-dimensional XXZ Hamiltonian \cite{Bonner1964,Yang1966}
\begin{equation} \label{eq: xxz_ham}
H 
= \sum_{\braket{pq}} \Big[ \frac{1}{2} \: ( \SP{p} \SM{q} + \SM{p} \SP{q} ) 
+ \Delta \: \SZ{p} \SZ{q} \Big],
\end{equation}
where $\braket{pq}$ represents nearest neighbor indices and $\Delta$ is an anisotropy parameter.
The XXZ Hamiltonian in \Eq{\ref{eq: xxz_ham}} is exactly solvable by Bethe ansatz. \cite{Takahashi1971,Gaudin1971} 
Still, traditional methods based on a single spin configuration state struggle to describe the strong correlation associated with the $| \Delta | \lesssim 1$ regime. \cite{Bishop1996,Bishop2000}
Although AGP is exact at $\Delta = -1$ \cite{RubioGarcia2019} and highly accurate for $\Delta < -1$, it is fairly poor for $\Delta > - 1$. 
It predicts an unphysical first-order phase transition around $\Delta = 1$, even in finite systems where there should be no such phase transition. \cite{Liu2023}
\Fig{\ref{fig: en_xxz_m16n8}} shows that the BTS energy curve remains smooth near $\Delta = 1$, and it qualitatively captures the exact energy curve for the $| \Delta | \lesssim 1$ regime of the XXZ model. 

We also compare the accuracy of reduced density matrices of BTS for the one-dimensional XXZ Hamiltonian.
Let us define the $S_z$-$S_z$ correlation function as 
\begin{equation} \label{eq: sz_sz_correlation}
F_p 
= \frac{1}{m} \: \sum_{q = 1}^m \: \braket{\SZ{q} \: \SZ{q + p}},
\end{equation}
where $0 \leq p \leq m - 1$ represents the distance between the spin sites, and $q + p$ takes into account the structure of a periodic boundary condition when applicable. 
We observe that BTS is as accurate as AGP or better than AGP in approximating the exact correlation functions for the one-dimensional XXZ Hamiltonian, and the difference between AGP and BTS is more visible around $\Delta = 1$, where methods based on AGP also have larger energy errors. \cite{Liu2023} 
\Fig{\ref{fig: spin_xxz_pbc_m16n8}} shows that BTS qualitatively reproduces the exact $S_z$-$S_z$ correlation functions near $\Delta = 1$. 
The interesting structure of correlation function values for AGP in \Fig{\ref{fig: spin_xxz_pbc_m16n8}} is due to its elementary symmetric polynomial structure, \cite{Khamoshi2019,LCAGP2021}
as expressed in \Eq{\ref{eq: agp_as_doci}}. 
Indeed, the left panel of \Fig{\ref{fig: spin_xxz_pbc_m16n8}} corresponds to the bimodal extreme AGP, where the AGP coefficients are proportional to $\pm 1$. \cite{Liu2023}
The binary tree polynomial structure of BTS generalizes AGP, thus achieving a more accurate spin correlation. 
However, BTS still retains some symmetric structure inherited from AGP, as evident from the BTS correlation function values shown in \Fig{\ref{fig: spin_xxz_pbc_m16n8}}.

\subsection{Size-consistency} \label{sec: size_consistency}


\begin{figure}[b]
\centering
\includegraphics[width=0.9\columnwidth]{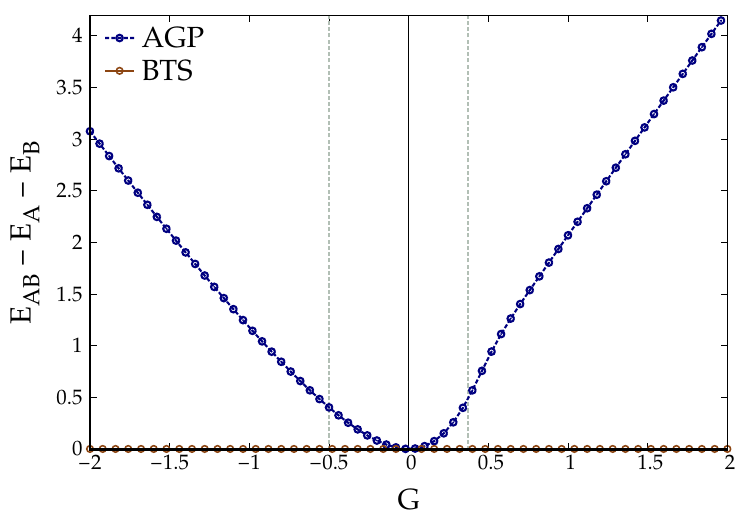}
\caption{
    Size-consistency errors corresponding to variational methods for a system composed of two identical non-interacting Hamiltonians. 
    Each of the Hamiltonians $H_A$ and $H_B$ are defined to be the pairing Hamiltonian of \Eq{\ref{eq: pairing_ham}} with $m = 8$ and $n = 4$.
    Each of the pairing Hamiltonian has critical $G$ values $G_c = - 0.5$ for the repulsive regime and $G_c \sim 0.3709$ for the attractive regime, represented by the two vertical grey dotted lines. 
    The energies $E_A$ and $E_B$ represent variationally optimal energies for the constituents $A$ and $B$, respectively, whereas $E_{AB}$ is the variationally optimal energy for the composite system. 
    The variational methods are AGP and BTS.
}
\label{fig: size_consistency}
\end{figure}


A desirable feature for any ansatz in quantum chemistry is size-consistency, which refers to the premise that the total energy of a system composed of non-interacting parts, when calculated as a unified entity, must coincide with the sum of the energies derived from each part independently. \cite{Pople1976} 
A size-consistent model is crucial when studying dissociation problems such as molecular bond-breaking. \cite{BartlettBook} 

Projective mean-field states such as AGP do not possess size-consistency beyond that of the Slater determinant, which it encompasses as a limit.
Specifically, if $\ket{\psi_A}$ and $\ket{\psi_B}$ are arbitrary AGP states for systems $A$ and $B$, respectively, then the tensor product $\ket{\psi_A} \otimes \ket{\psi_B}$ does not generally represent an AGP state corresponding to the entire system. 
It has been shown that the loss of size-consistency due to AGP can be recovered for a specific correlator ansatz based on the AGP reference. \cite{Neuscamman2012} 

Size-consistency for variational methods is often associated with a product state ansatz. 
In the context of BTS, the property of size consistency arises from the fact that the tensor product of two arbitrary BTS constitutes a BTS for the combined system. 
Specifically, consider the bipartition of a system into subsystems $A$ and $B$, and $\ket{\text{BTS}}_A$ and $\ket{\text{BTS}}_B$ are arbitrary BTS for the subsystems $A$ and $B$, respectively. 
Then the tensor product of these states constitutes a BTS for the entire system
\begin{equation} \label{eq: size_consistency_bts}
\ket{\text{BTS}}_{AB} 
= \ket{\text{BTS}}_A \otimes \ket{\text{BTS}}_B,
\end{equation}
where $\otimes$ represents a tensor product.

We numerically demonstrate the size-consistency of BTS, i.e., \Eq{\ref{eq: size_consistency_bts}}, by considering the dissociation limit of the pairing Hamiltonian in \Eq{\ref{eq: pairing_ham}}. 
We will focus on two distinct systems, $A$ and $B$, which are non-interacting and are collectively described by the following Hamiltonian: 
\begin{equation}
H 
= H_A + H_B,
\end{equation}
where $H_A$ and $H_B$ are identical pairing Hamiltonians, each encompassing eight paired levels. 
We focus on the half-filling case, where the ground state of $H$ should be a tensor product of the ground states of $H_A$ and $H_B$, each containing four pairs. 
The composite system represented by $H$ has sixteen paired levels, and we order them such that the levels of $H_A$ and $H_B$ belong to two different clusters, i.e., $\{ 1_A, \cdots, 8_A, 1_B, \cdots, 8_B \}$.
The clustered ordering of the paired levels is crucial for BTS to achieve size-consistency since BTS does not assume a symmetric decomposition of the DOCI coefficients.

We find the variationally optimal energies of systems $A$ and $B$ in isolation and for systems $A$ and $B$ combined, employing both AGP and BTS methods. 
\Fig{\ref{fig: size_consistency}} demonstrates that for the BTS method, the energies of the combined system $H_A + H_B$ are precisely the same as the summations of the energies for individual systems $A$ and $B$, thereby satisfying the condition of size-consistency. 
Conversely, AGP shows significant size-consistency errors in \Fig{\ref{fig: size_consistency}}, especially for large $G$ values of the pairing Hamiltonian. 


\section{Correlating BTS} \label{sec: correlation}

We have shown above that BTS can capture a significant part of the correlation energy.
Here, we discuss two possible routes for capturing the remaining correlations beyond single BTS. 

\subsection{Jastrow coupled cluster} \label{sec: jcc_bts}


\begin{figure}[b]
\centering
\includegraphics[width=0.9\columnwidth]{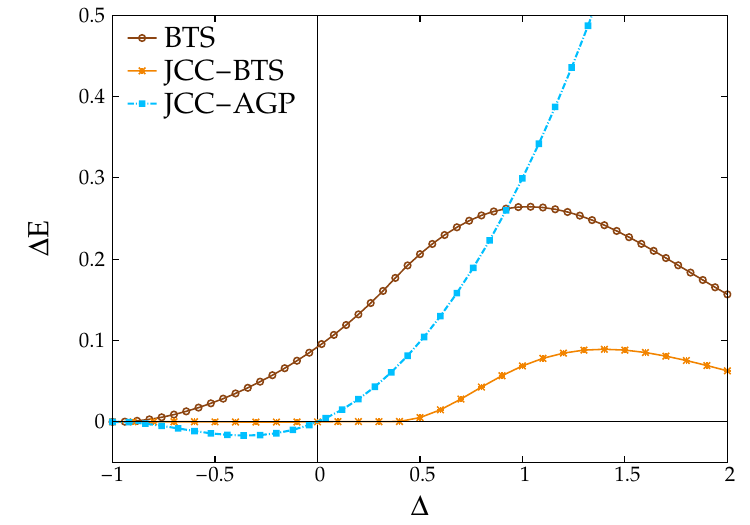}
\caption{ 
    Energy errors ($E_{\text{method}} - E_{\text{exact}}$) for the one-dimensional XXZ Hamiltonian with sixteen sites, $S_z = 0$, and open boundary conditions.
    The methods are BTS, J$_2$-CC on AGP, and J$_2$-CC on BTS.
}
\label{fig: en_bts_cc_xxz}
\end{figure}


\begin{figure}[t]
\centering
\includegraphics[width=0.9\columnwidth]{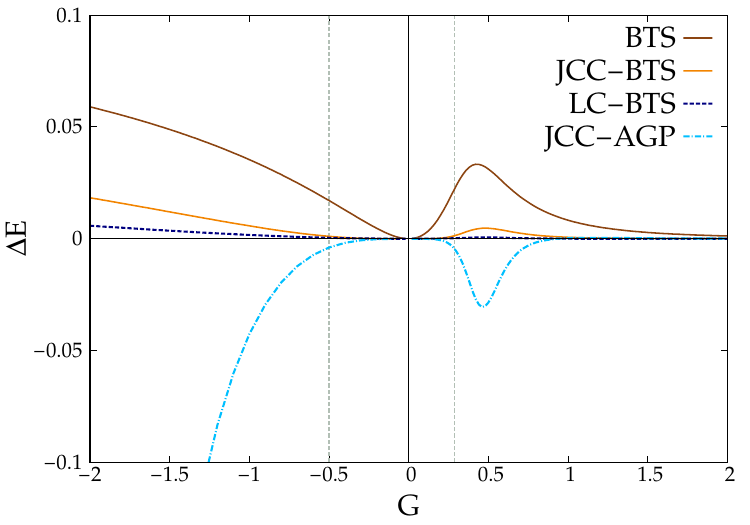}
\caption{
    Energy errors ($E_{\text{method}} - E_{\text{exact}}$) for the pairing Hamiltonian with $m = 16$ and $n = 8$.
    The critical $G$ values $G_c = - 0.5$ for the repulsive regime and $G_c \sim 0.2866$ for the attractive regime are represented by the two vertical grey dotted lines.
    The methods are BTS, J$_2$-CC on AGP, J$_2$-CC on BTS, and LC-BTS with two BT states. 
}
\label{fig: en_bts_corr_rbcs}
\end{figure}


The coupled cluster method, which uses an exponential correlator consisting of particle-hole excitations, is a popular route toward correlating Slater determinants. \cite{HelgakerBook,BartlettBook}
It has been shown that an efficient coupled cluster-like method for AGP \cite{KhamoshiCC2021} can be formulated with the two-body Hilbert space Jastrow correlator \cite{Neuscamman2013}
\begin{equation} \label{eq: j2_op}
J_2 
= \frac{1}{4} \sum_{1 \leq p < q \leq m} \alpha_{pq} \: \PN{p} \PN{q},   
\end{equation}
where $\alpha_{pq} = \alpha_{qp}$ are the correlator amplitudes. 
This Jastrow coupled cluster (JCC) formulation is the seniority-zero version of the Lie algebraic similarity transformation theory, \cite{WahlenStrothman2015,WahlenStrothman2016} 
and we here apply it to BTS.

Let us assume that DOCI can be approximated as 
\begin{equation}
\ket{\Psi}
= e^{J_2} \kBTS{m}{n},   
\end{equation}
and insert this into the Schr{\"o}dinger equation such that 
\begin{equation}
\bar{H} \ket{\Psi}
= E \kBTS{m}{n},   
\end{equation}
where $\bar{H} = e^{- J_2} H e^{J_2}$ is a similarity transformed Hamiltonian.
We extract the energy and correlator amplitudes $\{ \alpha_{pq} \}$ from
\begin{subequations} \label{eq: jcc_solve}
\begin{align}    
E 
&= \braket{ \BTP{m}{n} | \bar{H} | \BTP{m}{n} },  
\\
0 
&= \braket{ \BTP{m}{n} | \: \PN{p} \PN{q} \: ( \bar{H} - E) \: | \BTP{m}{n}}, \quad \forall \: p < q  
\end{align}
\end{subequations}
for a normalized $\kBTS{m}{n}$.
The computational bottleneck of \Eq{\ref{eq: jcc_solve}} is the expectation values
$\braket{\PN{p} \PN{q} \bar{H}}$ and $\braket{\PN{p} \PN{q}}$ for a given BTS reference, which we have chosen to be the variationally optimal single BTS.
Any $\braket{\PN{p} \PN{q}}$ element can be computed with $\ComCom{m n}$ complexity following \Appx{\ref{app: bts_rdms}}. 
We simplify 
$ \braket{\PN{p} \PN{q} \: \bar{H}} $ in \Appx {\ref{app: jcc_computation}}. \cite{KhamoshiCC2021}
The overall computational complexity of J$_2$-CC on BTS is $\ComCom{m^6}$.

\Fig{\ref{fig: en_bts_cc_xxz}} shows that J$_2$-CC energies for the one-dimensional XXZ Hamiltonian with open boundary conditions are more accurate with a BTS reference state instead of AGP, and the errors of BTS are largely mitigated by the inclusion of the $J_2$ correlator. 
For periodic boundary conditions, on the other hand, we have suffered from convergence difficulties. 
\Fig{\ref{fig: en_bts_corr_rbcs}} shows that BTS-based J$_2$-CC is also superior to AGP-based J$_2$-CC for the pairing Hamiltonian, particularly for repulsive interactions.

\subsection{Linear combination of BT states} \label{sec: lc_bts}


The simplest variational approach to go beyond a single BTS may be a linear combination of $R$ binary tree states (LC-BTS)
\begin{equation} \label{eq: lc_bts}
\ket{\Psi}
= \sum_{\mu = 1}^R \: \lambda_\mu \ket{\mu},   
\end{equation}
where $\{ \mu \}$ represents a fixed set of non-orthogonal BT states.
The expansion coefficients $\{ \lambda_\mu \}$ can be optimized as a non-orthogonal configuration interaction (NOCI)\cite{Urban1969} by solving the generalized eigenvalue problem
\begin{equation} \label{eq: ghep}
\mathbf{H} \bm{\Lambda}    
= \mathbf{M} \bm{\Lambda} \mathbf{E},
\end{equation}
where \textbf{H} and \textbf{M} in \Eq{\ref{eq: ghep}} are the Hamiltonian and metric matrices
\begin{subequations}
\begin{align}
H_{\mu \nu}
&= \braket{\mu | H | \nu}, 
\\
M_{\mu \nu}
&= \braket{\mu | \nu}, 
\end{align}
\end{subequations}
and where \textbf{E} and $\bm{\Lambda}$ are the eigenvalues and eigenvectors. 
Following \Appx{\ref{app: bts_rdms}}, building the \textbf{H} and \textbf{M} matrices has $\ComCom{R^2 m^4}$ complexity where $R$ is the number of BT states included in the NOCI.

Rather than working with a fixed set of BT states in the LC-BTS expansion, optimizing these states variationally to get a BTS analog of resonating Hartree--Fock theory is possible. \cite{Fukutome1988} 
In this case, we can absorb the variational coefficients $\{ \lambda_\mu \}$ by working with unnormalized BT states. 
The energy becomes
\begin{equation} \label{eq: en_lc_bts}
E ( \bm{\eta} )
= \frac{ \sum_{\mu \nu} \: \braket{\mu| H |\mu} }{ \sum_{\mu \nu} \: \braket{\mu| \nu } },   
\end{equation}
and we minimize it with respect to the $\{ \bm{\eta}^\mu \}$ vectors defining the BT states. 
\Fig{\ref{fig: en_bts_corr_rbcs}} shows results for this resonating BTS approach for the pairing Hamiltonian. 
Our approach optimizes only two BT states with $\ComCom{m^4}$ complexity, but still obtain energies more accurate than those from J2-CC on BTS, as shown in \Fig{\ref{fig: en_bts_corr_rbcs}}. 
However, the variational optimization approach is a computationally non-trivial problem, and while numerical techniques exist to deal with this problem,\cite{Fukutome1988,Schmid1989,JimenezHoyos2013,Mahler2021} 
we have observed convergence issues when applying this optimized LC-BTS approach with more than two BT states.
We have also observed similar convergence issues for the LC-BTS applied to the XXZ Hamiltonian, even when optimizing the ansatz with only two BT states. 


\section{Discussion} \label{sec: final}

We have introduced a classical algorithm for the BTS ansatz and explored it for approximating the eigenstates of \su2 Hamiltonians for which DOCI is exact.
Specifically, we have benchmarked methods based on BTS on the ground states of two Hamiltonians, one representing paired fermions and the other a prototypical \spinhalf Heisenberg model. 

BTS can be understood as an approximation to the computationally intractable APIG method.
Computing a general APIG reduced density matrix for a system with $m$ paired levels and $n$ pairs involves a summation over ${m \choose n}$ terms, where each term is a function of permanents. 
BTS handles the permanent part by keeping only one term from the permanent expansion approximating the corresponding DOCI coefficients.
The summations over ${m \choose n}$ terms are efficiently handled due to its binary tree structure. 

The monomial design of the BTS ansatz was inspired by AGP, which it generalizes with moderately more computational complexity.
This paper demonstrates that BTS can be significantly more accurate than AGP in model systems with a similar computational cost.
Unlike AGP, the BTS ansatz allows size-consistency, which is highly desirable.
Although this paper focuses on introducing an efficient classical implementation of BTS, we have also discussed directions for going beyond a single BTS. 

Because most electronic Hamiltonians do not have seniority as a symmetry,
developing post-BTS methods beyond seniority-zero systems requires choosing a suitable correlator that can couple different seniority sectors. \cite{Khamoshi2023} 
Choosing the best pairing scheme for $\{ p, \bar{p} \}$ may also be relevant. \cite{Bytautas2011,Limacher2014,Stein2014} 
We leave implementing methods based on BTS for general fermionic systems to future developments.


\begin{acknowledgments}

This work was supported by the U.S. Department of Energy, Office of Basic Energy Sciences, Computational and Theoretical Chemistry Program under Award No. DE-FG02-09ER16053. 
G.E.S. acknowledges support as a Welch Foundation Chair (Grant No. C-0036).

\end{acknowledgments}


\appendix


\section{Binary tree polynomial} \label{app: btp}

We discuss the BTP of \Eq{\ref{eq: btp}} in more detail here.
We can safely drop the BTP subscript and superscript symbols for convenience, which also helps to express \Eq{\ref{eq: btp}} as a function of matrix blocks without introducing too many symbols 
\begin{subequations}
\begin{align}
\BTP{}{} (\mathbf{A}_{2:4}^{1:2})
&= A_2^1 A_3^2 + A_2^1 A_4^2 + A_3^1 A_4^2, 
\\
\BTP{}{} (\mathbf{A}_{1:4}^{2:5})
&= A_1^2 A_2^3 A_3^4 A_4^5,
\end{align}    
\end{subequations}
where the superscript and subscript indicate the rows and columns of the parent \textbf{A} matrix. 

Even though \Eq{\ref{eq: btp}} is a function of the matrix \textbf{A}, it may not contain all its elements. 
A matrix element $A_p^j$ appears in $\BTP{m}{n} (\mathbf{A})$ only if the index $j$ belongs to the set 
\begin{equation} \label{eq: btp_set}
\BSet{m n}{p} 
= \{ j \: | \: \mbox{max} (1, p + n - m) \leq j \leq \mbox{min} (p, n) \}, 
\end{equation}
which originates from the fact that each of the monomial terms in \Eq{\ref{eq: btp}} is a product of $n$ scalars.
The $\{ A_j^p \}$ elements that do not appear in \Eq{\ref{eq: btp}} are irrelevant for our purposes, and we simply assume these to be zero.
The number of non-zero \textbf{A} matrix elements is 
\begin{equation} \label{eq: btp_dim}
d
= \sum_{p = 1}^m \: | \BSet{m n}{p} | 
\leq m \times n
\end{equation}
since the cardinality of $\BSet{m n}{p}$ can not be larger than $n$.

\subsection{Recursion relations} \label{app: btp_recursions}


\begin{figure}[t]
\centering
\includegraphics[width=0.9\columnwidth]{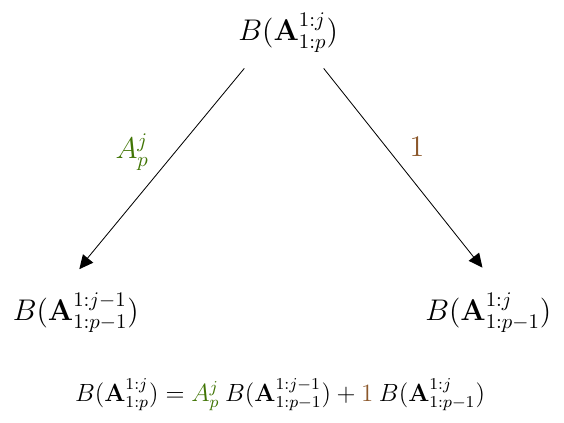}
\caption{
    Visualization of the terminal BTP recursion relation of \Eq{\ref{eq: btp_terminal_top}} with each node representing a BTP. 
    The rule is to multiply each element in a branch to the BTP corresponding to the node below.
    Then, summing over the two branch-weighted children BTPs equals their parent BTP.
    Similarly, each of the children nodes can have its branches and so on, which would create a binary tree. 
}
\label{fig: btp_terminal_recursion}
\end{figure}


BTP of \Eq{\ref{eq: btp}} can be computed recursively from 
\begin{subequations} \label{eq: btp_terminal_recursion}
\begin{align}
\BTP{}{} (\mathbf{A}_{1:p}^{1:j}) \label{eq: btp_terminal_top}  
&= A_p^j \: \BTP{}{} (\mathbf{A}_{1:p-1}^{1:j-1})   
+ \BTP{}{} (\mathbf{A}_{1:p-1}^{1:j}),   
\\
\BTP{}{} (\mathbf{A}_{p:m}^{j:n}) \label{eq: btp_terminal_bottom}   
&= A_p^j \: \BTP{}{} (\mathbf{A}_{p+1:m}^{j+1:n})   
+ \BTP{}{} (\mathbf{A}_{p+1:m}^{j:n}),
\end{align}
\end{subequations}
where $1 \leq j \leq n$ and $1 \leq p \leq m$ are defined based on a parent $\BTP{m}{n} (\mathbf{A})$ polynomial. 
We call \Eq{\ref{eq: btp_terminal_recursion}} the \textit{terminal} recursion relation of a BTP.
The binary tree structure of \Eq{\ref{eq: btp_terminal_top}} is illustrated in \Fig{\ref{fig: btp_terminal_recursion}}.
The terminal recursion relation of \Eq{\ref{eq: btp_terminal_recursion}} allows recursive computations of BTPs of all possible lower order and degrees before computing its \textit{root} polynomial.
As a result, any $\BTP{m}{n}$ polynomial can be computed with $\ComCom{m n}$ computational complexity.
We provide more details on computing BTP below.

A general recursion relation also exists for any BTP that goes beyond a binary tree 
\begin{align} \label{eq: btp_general_recursion}
\BTP{m}{n} (\mathbf{A})
&= \sum_{j \in \BSet{m n}{p}} A_p^j \: 
\BTP{}{} (\mathbf{A}_{1:p-1}^{1:j-1}) \: 
\BTP{}{} (\mathbf{A}_{p+1:m}^{j+1:n}) \nonumber 
\\
&+ \BTP{m - 1}{n} (\mathbf{A}_{- p}), 
\quad \forall \: p,   
\end{align}
where the set $\BSet{m n}{p}$ is defined in \Eq{\ref{eq: btp_set}} and $\mathbf{A}_{- p}$ represents a reduced matrix constructed by removing the $p\textsuperscript{th}$ column of its parent \textbf{A} matrix.
We prove \Eq{\ref{eq: btp_general_recursion}} in \Appx{\ref{app: btp_general_recursion}}. 

\subsection{BTP computation} \label{app: btp_computation}

The BTP computation algorithm here is an extension of the algorithm for computing BTP in \Reference{\citenum{ESPQC2023}} and shown in \Alg{\ref{alg: sum_btp}}. 
\Alg{\ref{alg: sum_btp}} can be regarded as a direct application of \Eq{\ref{eq: btp_terminal_recursion}} and indicates that computing $\BTP{m}{n}$ has $\ComCom{m n}$ computational complexity and comes with free computation of all BTPs with lower orders. 
Thus, \Alg{\ref{alg: sum_btp}} provides computation of  
$ \{ \BTP{}{} (\mathbf{A}_{1:p}^{1:j}) \} $ with $1 \leq j \leq n$ and $1 \leq p \leq m$, where we have applied the partition of \Eq{\ref{eq: btp_terminal_top}}.


\begin{algorithm}[h]
\SetAlgoNoLine    
\vskip1ex
  \KwIn{ Matrix $\mathbf{A}_{n \times m}$ with $n \leq m$. }
  \KwOut{ Matrix $\mathbf{B}_{n \times m}$ with $\BTP{p}{j} = \BTP{}{} (\mathbf{A}_{1:p}^{1:j})$. } 
  $ \BTP{p}{0} = 1, \quad 1 \leq p \leq m - 1; $ 
  \\
  $ \BTP{p}{j} = 0, \quad p < j; $
  \\
  $ \BTP{1}{1} = A_1^1; $
  \\
  \For{$p = 2:m$}{
  \For{$j = max(1, p + n - m):min(p, n)$}{
     \qquad $ \BTP{p}{j} 
            = A_p^j \: \BTP{p - 1}{j - 1}  + \BTP{p - 1}{j} $; 
  }}
  \Return{$\mathbf{B}_{1:m}^{1:n}$}
\caption{ 
    Computation of BTPs.
}
\label{alg: sum_btp}
\end{algorithm}


We can also compute 
$ \{ \BTP{}{} (\mathbf{A}_{p:m}^{j:n}) \} $ with $1 \leq j \leq n$ and $1 \leq p \leq m$ by applying the partition of \Eq{\ref{eq: btp_terminal_bottom}} in \Alg{\ref{alg: sum_btp}}.
However, another simple modification is possible.
Notice that the BTP expression of \Eq{\ref{eq: btp}} is invariant to the matrix transformation 
$ \bar{A}_p^j = A_{m - p + 1}^{n - j + 1} $.
Thus, the computation of 
$ \{ \BTP{}{} (\mathbf{A}_{p:m}^{j:n}) \} $
is equivalent to computing 
$ \{ \BTP{}{} (\mathbf{\bar{A}}_{1:p}^{1:j}) \} $ 
using \Alg{\ref{alg: sum_btp}}. 

\subsection{Proof of the general BTP recursion} \label{app: btp_general_recursion}

Let us partition the BTP $\BTP{m}{n}$ from \Eq{\ref{eq: btp}} based on the presence of $\mathbf{A}_p$ column elements 
\begin{equation} \label{eq: btp_simple_partition}
\BTP{m}{n} (\mathbf{A})
= \BTP{m - 1}{n} (\mathbf{A}_{- p}) 
+ \chi_p,
\end{equation}
where $\chi_p$ is the polynomial part with the terms of $\BTP{m}{n}$ that contain at least one of the elements from $\mathbf{A}_p$. 
We also know that the elements $\{ A_p^j \}$ which are present in $\chi_p$ depend on the set $\BSet{m n}{p}$ from \Eq{\ref{eq: btp_set}}. 
Based on the discussion so far, $\chi_p$ must have the functional form 
\begin{align} 
\chi_p 
&= \sum_{j \in \BSet{m n}{p}} \Big( 
\SumOp{p - 1}{j - 1} A_{p_1}^1 \cdots A_{p_{j - 1}}^{j - 1} \Big) \nonumber 
\\
&\quad \times \: A_p^j \: \Big( 
\sum_{p + 1 \leq p_{j + 1} < \cdots < p_n \leq m} A_{p_{j + 1}}^{j + 1} \cdots A_{p_n}^n \Big). 
\end{align} 
It is clear from the definition of BTP in \Eq{\ref{eq: btp}} that 
\begin{equation} \label{eq: btp_contains_element}
\chi_p
= \sum_{j \in \BSet{m n}{p}} A_p^j \: 
\BTP{}{} (\mathbf{A}_{1:p-1}^{1:j-1}) \: 
\BTP{}{} (\mathbf{A}_{p+1:m}^{j+1:n}),
\end{equation}
which proves \Eq{\ref{eq: btp_general_recursion}}.



\section{APIG and BTS} \label{app: apig_and_bts}

It may be tempting to think that APIG reduces to BTS by following the constraint 
\begin{align} \label{eq: bts_constraint}
\eta_p^j 
= 0, \quad \text{if} \: j \notin \BSet{m n}{p},
\end{align}
where the set $\BSet{m n}{p}$ is defined in \Eq{\ref{eq: btp_set}}.
However, this is not necessarily the case, as shown below with an example of a system with five paired orbitals and three pairs:
\begin{widetext}
\begin{align} \label{eq: a43_example}
\ket{\text{APIG}}
&= \Big( \eta_1^1 \eta_2^2 \eta_3^3 
       + \eta_1^1 \eta_3^2 \eta_2^3 
       + \eta_2^1 \eta_1^2 \eta_3^3 
       + \eta_2^1 \eta_3^2 \eta_1^3 
       + \eta_3^1 \eta_1^2 \eta_2^3 
       + \eta_3^1 \eta_2^2 \eta_1^3 
    \Big) \ket{11100} \nonumber 
\\
&+  \Big( \eta_1^1 \eta_2^2 \eta_4^3 
       + \eta_1^1 \eta_4^2 \eta_2^3 
       + \eta_2^1 \eta_1^2 \eta_4^3 
       + \eta_2^1 \eta_4^2 \eta_1^3 
       + \eta_4^1 \eta_1^2 \eta_2^3 
       + \eta_4^1 \eta_2^2 \eta_1^3 
    \Big) \ket{11010} \nonumber
\\
&+  \Big( \eta_1^1 \eta_2^2 \eta_5^3 
       + \eta_1^1 \eta_5^2 \eta_2^3 
       + \eta_2^1 \eta_1^2 \eta_5^3 
       + \eta_2^1 \eta_5^2 \eta_1^3 
       + \eta_5^1 \eta_1^2 \eta_2^3 
       + \eta_5^1 \eta_2^2 \eta_1^3 
    \Big) \ket{11001} \nonumber
\\
&+  \Big( \eta_1^1 \eta_3^2 \eta_4^3 
       + \eta_1^1 \eta_4^2 \eta_3^3 
       + \eta_3^1 \eta_1^2 \eta_4^3 
       + \eta_3^1 \eta_4^2 \eta_1^3 
       + \eta_4^1 \eta_1^2 \eta_3^3 
       + \eta_4^1 \eta_3^2 \eta_1^3 
    \Big) \ket{10110} \nonumber
\\
&+  \Big( \eta_1^1 \eta_3^2 \eta_5^3 
       + \eta_1^1 \eta_5^2 \eta_3^3 
       + \eta_3^1 \eta_1^2 \eta_5^3 
       + \eta_3^1 \eta_5^2 \eta_1^3 
       + \eta_5^1 \eta_1^2 \eta_3^3 
       + \eta_5^1 \eta_3^2 \eta_1^3 
    \Big) \ket{10101} \nonumber
\\
&+  \Big( \eta_1^1 \eta_4^2 \eta_5^3 
       + \eta_1^1 \eta_5^2 \eta_4^3 
       + \eta_4^1 \eta_1^2 \eta_5^3 
       + \eta_4^1 \eta_5^2 \eta_1^3 
       + \eta_5^1 \eta_1^2 \eta_4^3 
       + \eta_5^1 \eta_4^2 \eta_1^3 
    \Big) \ket{10011} \nonumber
\\
&+  \Big( \eta_2^1 \eta_3^2 \eta_4^3 
       + \eta_2^1 \eta_4^2 \eta_3^3 
       + \eta_3^1 \eta_2^2 \eta_4^3 
       + \eta_3^1 \eta_4^2 \eta_2^3 
       + \eta_4^1 \eta_2^2 \eta_3^3 
       + \eta_4^1 \eta_3^2 \eta_2^3 
    \Big) \ket{01110} \nonumber
\\
&+  \Big( \eta_2^1 \eta_3^2 \eta_5^3 
       + \eta_2^1 \eta_5^2 \eta_3^3 
       + \eta_3^1 \eta_2^2 \eta_5^3 
       + \eta_3^1 \eta_5^2 \eta_2^3 
       + \eta_5^1 \eta_2^2 \eta_3^3 
       + \eta_5^1 \eta_3^2 \eta_2^3 
    \Big) \ket{01101} \nonumber
\\
&+  \Big( \eta_2^1 \eta_4^2 \eta_5^3 
       + \eta_2^1 \eta_5^2 \eta_4^3 
       + \eta_4^1 \eta_2^2 \eta_5^3 
       + \eta_4^1 \eta_5^2 \eta_2^3 
       + \eta_5^1 \eta_2^2 \eta_4^3 
       + \eta_5^1 \eta_4^2 \eta_2^3 
    \Big) \ket{01011} \nonumber
\\
&+  \Big( \eta_3^1 \eta_4^2 \eta_5^3 
       + \eta_3^1 \eta_5^2 \eta_4^3 
       + \eta_4^1 \eta_3^2 \eta_5^3 
       + \eta_4^1 \eta_5^2 \eta_3^3 
       + \eta_5^1 \eta_3^2 \eta_4^3 
       + \eta_5^1 \eta_4^2 \eta_3^3 
    \Big) \ket{00111},
\end{align}
\end{widetext}
where the strings are the paired Slater determinants with $0$ and $1$ representing empty and occupied paired orbitals.
Applying \Eq{\ref{eq: bts_constraint}} reduces \Eq{\ref{eq: a43_example}} to
\begin{align} \label{eq: a43_final}
\ket{\text{APIG}}
&= \eta_1^1 \eta_2^2 \eta_3^3 \ket{11100} 
+ \eta_1^1 \eta_2^2 \eta_4^3 \ket{11010} 
+ \eta_1^1 \eta_2^2 \eta_5^3 \ket{11001} \nonumber
\\ 
&+ \Big( \eta_1^1 \eta_3^2 \eta_4^3 
+ \eta_1^1 \eta_4^2 \eta_3^3 \Big) \ket{10110} \nonumber 
+ \eta_1^1 \eta_3^2 \eta_5^3 \ket{10101} \nonumber
\\
&+ \eta_1^1 \eta_4^2 \eta_5^3 \ket{10011}  
+ \Big( \eta_2^1 \eta_3^2 \eta_4^3 
+ \eta_2^1 \eta_4^2 \eta_3^3  
+ \eta_3^1 \eta_2^2 \eta_4^3 \Big) \ket{01110} \nonumber 
\\
&+ \Big( \eta_2^1 \eta_3^2 \eta_5^3
+ \eta_3^1 \eta_2^2 \eta_5^3 \Big) \ket{01101} \nonumber  
\\
&+ \eta_2^1 \eta_4^2 \eta_5^3 \ket{01011} 
+ \eta_3^1 \eta_4^2 \eta_5^3 \ket{00111}, 
\end{align}
which is not a BTS. 
The state in \Eq{\ref{eq: a43_final}} reduces to BTS only when we eliminate the terms whose subscript and superscript indices in $\{ \eta_p^j \}$ are not in increasing order. 

\section{BTS transformations} \label{app: bts_transformation}

A general BTS recursion relation can be derived by direct application of \Eq{\ref{eq: btp_general_recursion}} to BTS 
\begin{align} \label{eq: bts_general_recursion}
\kBTS{m}{n}
&= \sum_{j \in \BSet{m n}{p}} \eta_p^j \PC{p} \: \BTP{}{} (\mathbf{A}_{1:p-1}^{1:j-1}) \: 
\BTP{}{} (\mathbf{A}_{p+1:m}^{j+1:n}) \ket{-} \nonumber 
\\
&+ \kBTS{m - 1}{n}_{- p}, \quad \forall \: p,  
\end{align}
where any $ A_p^j = \eta_p^j \PC{p} $ and $- p$ indicates removal of the $p\textsuperscript{th}$ paired orbital from the corresponding BTS expansion. 
The set $\BSet{mn}{p}$ in \Eq{\ref{eq: bts_general_recursion}} is defined in \Eq{\ref{eq: btp_set}}. 
\Eq{\ref{eq: bts_general_recursion}} allows us to write 
\begin{equation} \label{eq: pa_on_bts} 
\PA{p} \kBTS{m}{n}
= \sum_{j \: \in \: \BSet{m n}{p}} \eta_p^j \: \BTP{}{} (\mathbf{A}_{1:p-1}^{1:j-1}) \: 
\BTP{}{} (\mathbf{A}_{p+1:m}^{j+1:n}) \ket{-}.
\end{equation}
Let us also define a \textit{shifted} paired orbital number operator as $\SPN{p} = \PN{p} + \beta$, where $\beta$ is an arbitrary scalar. 
\Eq{\ref{eq: bts_general_recursion}} indicates that the $\SPN{p}$ operator acts on BTS to create another BTS 
\begin{equation} \label{eq: nbar_on_bts}
\kredBTS{m n}{p}
= \frac{1}{\beta} \: \SPN{p} \kBTS{m}{n},
\end{equation}
where the coefficients of the modified BTS $\kredBTS{m n}{p}$ are 
\begin{equation}
\bar{\eta}_p^j
= \big( 1 + \frac{2}{\beta} \big) \: \eta_p^j, \quad \forall \: j.   
\end{equation}
As shown in \Reference{\citenum{LCAGP2021}}, the exponential of the one-body Jastrow operator 
\begin{equation}
J_1
= \frac{1}{2} \: \sum_{p = 1}^m \: \alpha_p \: \PN{p}   
\end{equation}
can be expanded as a product of shifted paired orbital number operators 
\begin{equation}
e^{J_1}
\propto \prod_{p = 1}^m \: 
\Big( \PN{p} + \frac{2}{e^{\alpha_p} - 1} \Big),   
\end{equation}
where we have ignored an irrelevant global scalar factor.
Thus, the operator $e^{J_1}$ on BTS simply transforms it into another BTS. \cite{ESPQC2023}


\section{Reduced density matrices} \label{app: bts_rdms}

We discuss computing transition reduced density matrix (RDM) elements between different binary tree states. 
Any transition RDM element has the form $\braket{\mu| \cdots |\nu}$ with an arbitrary string of operators in the middle. 
The state $\ket{\mu}$ represents the $\mu\textsuperscript{th}$ state of a set of BT states with its coefficients defined by the matrix $\bm{\eta}^\mu$. 

\subsection{Case I: Diagonal elements}

The relation $\PN{p} = 2 \: \PC{p} \PA{p}$ is true when $\PN{p}$ act on a state of the form in \Eq{\ref{eq: doci}}.
It is then straightforward to show from \Eq{\ref{eq: pa_on_bts}} that 
\begin{equation} \label{eq: z11_bts} 
\frac{1}{2} \braket{\PN{p}}
= \sum_{j \: \in \: \BSet{m n}{p}} 
( \eta_p^{\mu j} )^* \: \eta_p^{\nu j} \: \BTP{}{} (\mathbf{X}_{1:p-1}^{1:j-1}) \: 
\BTP{}{} (\mathbf{X}_{p+1:m}^{j+1:n}), 
\end{equation}
where $\braket{\PN{p}}$ represents $\braket{\mu| \PN{p} |\nu}$ and $X_p^j = ( \eta_p^{\mu j} )^* \: \eta_p^{\nu j}$. 
Following \Appx{\ref{app: btp_computation}}, after storing all the BTP intermediates in \Eq{\ref{eq: z11_bts}} with $\ComCom{m n}$ space, any $ \braket{\PN{p}} $ is computable with $\ComCom{n}$ complexity.

\subsection{Case II: Off-diagonal elements}

We discuss the computation of 
$ \braket{\BTP{m}{n}| \PC{p} \PA{q} |\BTP{m}{n}} $, where $p$ and $q$ are different.
We assume the bra and ket states to be the same BTS for the sake of simplicity, but the extension to the general case is straightforward following \Eq{\ref{eq: z11_bts}}.
Let us also assume $p < q < m$ since the $p > q$ case can be similarly derived and $q = m$ is trivial due to \Eq{\ref{eq: bts_terminal_recursion}}.
We extend \Eq{\ref{eq: bts_terminal_recursion}} to arrive at 
\begin{subequations} \label{eq: pa_bts_cases}
\begin{align}
\PA{q} \kBTS{r}{j} 
&= \eta_q^j \kBTS{q - 1}{j - 1}, \quad \forall \: q = r, 
\\
&= 0, \quad \forall \: q > r, 
\\
&= \PA{q} \kBTS{r}{j}, \quad \forall \: q < r.
\end{align}   
\end{subequations}
Let us define two types of tensor elements
\begin{subequations} 
\begin{align}
V_{pq}^{rj}
&= \braket{\BTP{r}{j} | \PC{p} \PA{q} | \BTP{r}{j}},
\\    
W_p^{rj}
&= \braket{\BTP{r}{j - 1} | \PA{p} | \BTP{r}{j}},
\end{align}
\end{subequations}
where $1 \leq r \leq m$ and $1 \leq j \leq n$. 
We now apply \Eq{\ref{eq: bts_terminal_recursion}} and \Eq{\ref{eq: pa_bts_cases}} to $\braket{\PC{p} \PA{q}}$ to arrive at 
\begin{subequations} \label{eq: z02_int_recursion} 
\begin{align}
V_{pq}^{rj}
&= \eta_q^j \: ( W_p^{r - 1, j} )^*, \quad \forall \: q = r,
\\
&= | \eta_r^j |^2 \: V_{pq}^{r - 1, j - 1}
+ V_{pq}^{r - 1, j}, \quad \forall \: q < r.
\end{align}
\end{subequations} 
It can be similarly shown that
\begin{subequations} \label{eq: z01_int_recursion}  
\begin{align}
W_p^{rj}
&= \eta_p^j \braket{\BTP{j - 1}{p - 1}| \BTP{j - 1}{p - 1}}, \: \forall \: p = r,
\\
&= ( \eta_r^{j - 1} )^* \: \eta_r^j \: W_p^{r - 1, j - 1}
+ W_p^{r - 1, j}, \: \forall \: p < r.
\end{align}
\end{subequations} 
Both \Eq{\ref{eq: z02_int_recursion}} and \Eq{\ref{eq: z01_int_recursion}} have a binary tree structure similar to \Eq{\ref{eq: btp_terminal_recursion}}, which we utilize to design algorithms similar to \Alg{\ref{alg: sum_btp}}.


\begin{algorithm}[h]
\SetAlgoNoLine    
\vskip1ex
  \KwIn{ Matrix $\bm{\eta}_{n \times m}$ with $n \leq m$ and integer $p$ with $1 < p < m$. }
  \KwOut{ Scalar $ \braket{\BTP{m}{n - 1}| \PA{p} | \BTP{m}{n}} $. } 
  $ W_j^p = \eta_p^j \braket{\BTP{p - 1}{j - 1} | \BTP{p - 1}{j - 1}} $ using \Alg{\ref{alg: sum_btp}};
  \\
  \For{$r = p+1:m$}{
  \For{$j = max(1, r + n - m):min(r, n)$}{
     \qquad $ W_j^r
            = ( \eta_r^{j - 1} )^* \: \eta_r^j \: W_{j - 1}^{r - 1}
            + W_j^{r - 1} $; 
  }}
  \Return{$W_n^m$}
\caption{ 
    Computation of $ \braket{\BTP{m}{n - 1}| \PA{p} | \BTP{m}{n}} $.
}
\label{alg: z01_bts}
\end{algorithm}


\begin{algorithm}[h]
\SetAlgoNoLine    
\vskip1ex
  \KwIn{ Matrix $\bm{\eta}_{n \times m}$ with $n \leq m$ and integer $p$ and $q$ with $1 < p < q < m$. }
  \KwOut{ Scalar $ \braket{\BTP{m}{n}| \PC{p} \PA{q} | \BTP{m}{n}} $. } 
  $ V_j^q = \eta_q^j \braket{\BTP{q - 1}{j - 1} | \PA{p} | \BTP{q - 1}{j}}^* $ using \Alg{\ref{alg: z01_bts}};
  \\
  \For{$r = q+1:m$}{
  \For{$j = max(1, r + n - m):min(r, n)$}{
     \qquad $ V_j^r
            = | \eta_r^j |^2 \: V_{j - 1}^{r - 1}
            + V_j^{r - 1} $; 
  }}
  \Return{$V_n^m$}
\caption{ 
    Computation of $ \braket{\BTP{m}{n}| \PC{p} \PA{q} | \BTP{m}{n}} $.
}
\label{alg: z02_bts}
\end{algorithm}


We first obtain all BTS overlap intermediates with \Alg{\ref{alg: sum_btp}}  and store them in $\ComCom{m n}$ space. 
Then the $W_p^{rj}$ and $V_{pq}^{rj}$ tensor elements are computed with \Alg{\ref{alg: z01_bts}} and \Alg{\ref{alg: z02_bts}}, respectively.   
Overall, the computation of each of the elements
$ \braket{\BTP{m}{n}| \PC{p} \PA{q} |\BTP{m}{n}} $ has complexity $\ComCom{m n}$ with optimal storage of intermediates.

Algorithms for higher-order off-diagonal elements, e.g., 
$ \braket{\PC{p} \PC{q} \PA{r} \PA{s}} $ can be similarly designed by extending the recursion relation in \Eq{\ref{eq: pa_bts_cases}}. 

\subsection{Case III: Mixed elements}

Assuming $\braket{\mathcal{S}}$ is easily computable with $\mathcal{S}$ representing an operator string, any RDM element of the form 
$ \braket{ \PN{p_1} \cdots \PN{p_j} \: \mathcal{S} \PN{p_1} \cdots \PN{p_k} } $ 
can be efficiently computed due to \Eq{\ref{eq: nbar_on_bts}}.
Let us discuss with a simple example
\begin{align}
\braket{\PN{p} \PC{q} \PA{r} \PN{s}}   
&= \braket{\SPN{p} \PC{q} \PA{r} \SPN{s}} 
- \beta^* \braket{\PC{q} \PA{r} \PN{s}} 
\\
&- \beta \braket{\PN{p} \PC{q} \PA{r}} 
- |\beta|^2 \braket{\PC{q} \PA{r}}, 
\end{align}
where the lower-order RDM elements are 
\begin{align}
\braket{\PC{q} \PA{r} \PN{s}}   
&= \braket{\PC{q} \PA{r} \SPN{s}}   
- \beta \braket{\PC{q} \PA{r}},  
\\
\braket{\PN{p} \PC{q} \PA{r}} 
&= \braket{\SPN{p} \PC{q} \PA{r}} 
- \beta^* \braket{\PC{q} \PA{r}}. 
\end{align}
The RDM elements 
$ \braket{\SPN{p} \PC{q} \PA{r} \SPN{s}} $,
$ \braket{\PC{q} \PA{r} \SPN{s}} $, and  
$ \braket{\SPN{p} \PC{q} \PA{r}} $ are computable as $\braket{\PC{q} \PA{r}}$ corresponding to the BT states defined following \Eq{\ref{eq: nbar_on_bts}}.


\section{Coupled cluster residual} \label{app: jcc_computation}

We simplify 
$ \braket{\PN{p} \PN{q} \: \bar{H}} $ from \Eq{\ref{eq: jcc_solve}} here. 
It is easy to see that  
\begin{align}
\braket{\PN{p} \PN{q} \: \bar{H}}  
&= \braket{ ( \PN{p} + \beta ) \: ( \PN{q} + \beta ) \: \bar{H}} \nonumber 
\\
&- \beta \braket{ \PN{q} \: \bar{H}} 
- \beta \braket{ \PN{p} \: \bar{H}} 
- \beta^2 \braket{\bar{H}}, 
\end{align}
where $\beta$ is an arbitrary scalar. 
Similarly, the same approach leads to  
\begin{equation}
\braket{\PN{p} \: \bar{H}}  
= \braket{ ( \PN{p} + \beta ) \: \bar{H}} 
- \beta \braket{ \bar{H}}. 
\end{equation}
We show in \Appx{\ref{app: bts_transformation}} that an operator of the form $\PN{p} + \beta$ acting on BTS simply transforms it into another BTS. 
Thus, computing 
$ \braket{\PN{p} \PN{q} \: \bar{H}} $ involves BTS transition expectation value with respect to the $\bar{H}$ operator.

The similarity transformation of a Hamiltonian with certain orbital number operators can be resummed. \cite{WahlenStrothman2015,WahlenStrothman2016,KhamoshiCC2021}  
Specifically, it can be shown that 
\begin{subequations}
\begin{align}
e^{- J_2} \: \PC{p} \: e^{J_2} 
&= e^{- J_{1 p}} \: \PC{p} \: e^{- J_{1 p}},
\\    
e^{- J_2} \: \PA{p} \: e^{J_2} 
&= e^{J_{1 p}} \: \PA{p} \: e^{J_{1 p}},
\\    
e^{- J_2} \: \PN{p} \: e^{J_2} 
&= \PN{p},
\end{align}
\end{subequations}
where $J_{1 p}$ is a special one-body Jastrow operator dependent on its parent two-body Jastrow operator in \Eq{\ref{eq: j2_op}}
\begin{equation} \label{eq: j1p_op}
J_{1 p}  
= \frac{1}{4} \: \sum_{q = 1}^m \: \alpha_{pq} \: \PN{q},  
\end{equation}
and any $\alpha_{pp} = 0$.
Since any exponential of one-body Jastrow operator acting on BTS creates another BTS, computing 
$ \braket{\PN{p} \PN{q} \: \bar{H}} $ is equivalent to computing BTS Hamiltonian transition overlap between two different binary tree states. 

\section{BTS and angular momentum coupling} \label{app: CGtoBTS}


\begin{figure}[t]
\centering
\includegraphics[width=0.9\columnwidth]{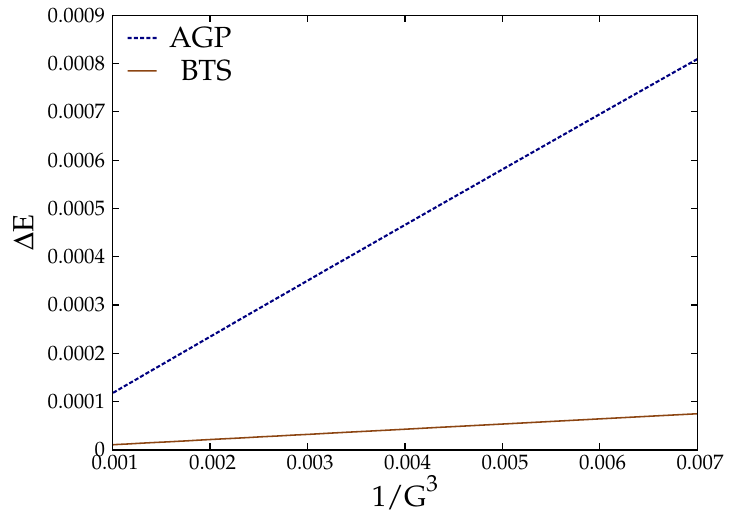}
\caption{   
   Energy errors ($E_{\text{method}} - E_{\text{exact}}$) for the pairing Hamiltonian at the attractive regime with small $1/G^3$ values.
   The system has $m = 16$ paired levels and $n = 8$ pairs.
}
\label{fig: en_bts_rbcs_attr_m16n8}
\end{figure}


We discuss how a state characterized by a fixed total spin $S$ and a projected spin $S_z$ can be associated with a BTS through an analogy between the angular momentum coupling equations and the BTS recursion relation of \Eq{\ref{eq: bts_terminal_recursion}}.
Consequently, we show the wavefunction given by optimized BTS is exact for the pairing Hamiltonian of \Eq{\ref{eq: pairing_ham}} for $G = \pm\infty$. Moreover, we summarize the asymptotic behaviors as $G \rightarrow \pm \infty$ and explain them using perturbation theory. 

As described in \Eq{\ref{eq: bts_terminal_recursion}}, the BTS recursion relation can be interpreted as an iterative coupling of the site $p$ and can be reformulated as follows:
\begin{equation} \label{eq: ite_BTS}
\kBTS{p}{j}
= \eta_p^j \kBTS{p-1}{j-1} \otimes \ket{\uparrow}_p 
+ \kBTS{p-1}{j} \otimes \ket{\downarrow}_p,
\end{equation}
where we employed the pair to \spinhalf mapping of \Eq{\ref{eq: pair_spin_mapping}}. 
The formula for angular momentum coupling is given as \cite{ShankarBook}
\begin{equation}
|JM\rangle 
= \sum_{\substack{ m_1 m_2 \\ m_1 + m_2 = M }} C_{j_1 m_1 j_2 m_2}^{JM} \ket{ j_1 m_1 } \otimes \ket{ j_2 m_2 },
\end{equation}
where $\{ C_{j_1 m_1 j_2 m_2}^{JM} \}$ are the Clebsch--Gordan(CG) coefficients. 
We focus on iteratively coupling a new \spinhalf site in this context. 
Consequently, we can confine our attention to cases where $j_2 = \frac{1}{2}$. With these considerations, we arrive at
\begin{align} \label{eq: ite_CG}
&\ket{ j_1 \pm \frac{1}{2}, M } 
= \pm \sqrt{\frac{1}{2} \left( 1 \pm \frac{M}{j_1 + \frac{1}{2}} \right)} 
\ket{ j_1, M - \frac{1}{2} } \otimes \ket{ \uparrow } \nonumber 
\\ 
&\hskip7ex 
+ \sqrt{\frac{1}{2} \left( 1 \mp \frac{M}{j_1 + \frac{1}{2}} \right)} \ket{ j_1, M+ \frac{1}{2} } \otimes \ket{ \downarrow }. 
\end{align} 
A state characterized by a fixed total spin $S$ and a projected spin $S_z$ can then be associated with a BTS by iteratively applying \Eq{\ref{eq: ite_BTS}} to \Eq{\ref{eq: ite_CG}}. 
Following \Reference{\citenum{ESPQC2023}}, we can apply the recursion relation for a normalized BTS to derive a closed-form relationship between the BTS and CG coefficients of such states.

When considering the pairing Hamiltonian, it is evident that the two-body interaction term dominates for both strongly repulsive $G = -\infty$ and attractive $G = +\infty$ regimes.
As a result, the Hamiltonian simplifies to $-G \: S^{+} S^{-}$, which represents a homogeneous long-range spin-spin interaction and can be reformulated as 
\begin{equation}
H 
= - G \: ( S^2 - S_z^2 + S_z ).
\end{equation}
Since we are focused on systems with a constant $S_z$, the energy must depend solely on $S$. 
Consequently, the eigenstates of the pairing Hamiltonian will possess a fixed value of $S$.

The value of $S$ for the ground state is contingent upon the sign of $G$. 
Specifically, as $G = -\infty$, the ground state assumes $S = 0$. 
Conversely, when $G = +\infty$, the ground state (which is of spin-AGP form) takes on $S=\frac{M}{2}$, representing the maximal $S$ attainable through coupling $M$ \spinhalf sites.
Since the ground state is an eigenstate of the total spin operator $S$ and the projected spin operator $S_z$ for $G = \pm\infty$, it can be represented by a BTS by mapping the CG coefficients to BTS coefficients. 
Consequently, BTS is energetically exact as $G = \pm\infty$, while AGP is exact only for $G = +\infty$.
The numerical validation of the preceding arguments, as illustrated in the right panel of \Fig{\ref{fig: en_bts_rbcs_repul_m16n8}} and \Fig{\ref{fig: en_bts_rbcs_attr_m16n8}}, reveals distinct scaling behaviors of energy error $\Delta E$. 
In the limit where $G \rightarrow + \infty$, we have observed both $\Delta E_{BTS}$ and $\Delta E_{AGP}$ to scale as $O(\frac{1}{G^3})$, albeit with different prefactors. 
On the other hand, we have observed that the $\Delta E_{BTS}$ follows a scaling of $O(\frac{1}{|G|})$ as $G \rightarrow - \infty$.

To explain the asymptotic behavior of the pairing Hamiltonian, we need to consider the situation where $|G|$ is large but still finite, which leads to the following form
\begin{subequations}
\begin{align}
    H 
    &= G \: \Big[ - \sum_{pq} S_p^{\dagger}S_q + \frac{1}{G} \sum_p \epsilon_p \: (2S^z_p+1) \Big]  
    \\
    &= G \: \Big[ - S^2 +S_z^2 + S_z 
    + \frac{1}{G} \sum_p \epsilon_p \: (2S^z_p+1) \Big].
\end{align}
\end{subequations}
Since we only consider states with $S_z = 0$, the Hamiltonian can be written, up to 
a constant, as
\begin{equation}
    H 
    = G \: ( -S^2 + \frac{1}{G} \sum_p 2 \epsilon_p \: S_p^z ).
\end{equation}
Let us first focus on the attractive regime $G \rightarrow +\infty $, where the Hamiltonian can be written as
\begin{equation}
    H = G \: (H_0 + \frac{1}{G} \: H_1), 
\end{equation}
where the zeroth-order Hamiltonian is 
$ H_0 = -S^2 $ with a unique ground state that has $S = M/2$ and $S_z = 0$ 
\begin{equation}
\ket{\text{eAGP}}
= \SumOp{m}{n} \SP{p_1} \cdots \SP{p_n} \ket{\downarrow \cdots \downarrow},
\end{equation}
which is known as extreme AGP. \cite{Liu2023}
The perturbation term 
$ H_1 = \sum_p 2 \epsilon_p \: S_p^z $ is an operator tensor of $S = 1$. 
By the spin selection rule, we have 
$ \braket{ \text{eAGP} | H_1 | \text{eAGP} } = 0 $. 
Moreover,  $H_1 \ket{\text{eAGP}}$ is a state with $S = M/2 - 1$, which is an eigenstate of $H_0$. 
The first-order wavefunction can then be simplified as
\begin{align}
    \ket{ \psi_1 } 
    &= \frac{1}{E_0 - H_0} \: H_1 \ket{\text{eAGP}} \nonumber
    \\
    &= \frac{1}{E_0 - E_1} H_1 \ket{\text{eAGP}},
\end{align}
where $E_0$ and $E_1$ are the eigenvalues of $H_0$ with respect to $S = M/2$ and $S = M/2 - 1$, respectively. 
The ground state wavefunction can then be written as
\begin{equation}
    \ket{ \psi } 
    = \ket{ \text{eAGP} } 
    + \frac{ \sum_p \: 2 \epsilon_p \: S^z_p }{ G ( E_0 - E_1 ) } \ket{ \text{eAGP} } 
    + O(1/G^2),
\end{equation}
which can be approximated by an AGP with an error of order $1/G^2$
\begin{equation}
    \ket{\psi } 
    = e^{J_1/G} \ket{\text{ eAGP }} 
    + O(1/G^2),
\end{equation}
where $J_1 = \frac{\sum_p 2\epsilon_p S^z_p}{E_0-E_1} $ and 
$ e^{J_1/G} \ket{\text{ eAGP }} $ is simply another AGP. \cite{KhamoshiCC2021}
From the above analysis, the wavefunction of AGP is correct up to $1/G$. Consequently, the energy is correct up to $1/G^3$ by Wigner's $2n + 1$ rule. \cite{BartlettBook}
BTS has the same asymptotic behavior in the attractive regime because it encompasses the AGP wavefunction. 

For the repulsive regime of the pairing Hamiltonian, where $G\rightarrow -\infty$, we write the Hamiltonian as
\begin{equation}
    H = |G| \: ( H_0 + \frac{1}{|G|} \: H_1),
\end{equation}
where the zeroth-order Hamiltonian is 
$ H_0 = S^2 $, its ground states are those with $S = 0, S_z = 0$ and has a high degree of degeneracy. 
By the selection rule, $H_1$ will lift the states with $S = 0$ to $S = 1$ and vice versa, resulting in a second-order effective Hamiltonian acting on the $S = 0$ subspace
\begin{equation}
    H_{\text{eff}} 
    = P_{S=0} \: H_1 \: \frac{1}{E_0 - E_1} \: H_1 \: P_{S=0},
\end{equation}
where $P_{S=0}$ is the projection operator onto the $S = 0$ subspace, and $E_0$ and  $E_1$ are the eigenvalues of $H_0$ with respect to $S = 0$ and $S = 1$. 
Since $H_{\text{eff}}$ is a second-order term, the energy of BTS is correct up to $1/|G|$. 
However, the exact ground state for $G \rightarrow -\infty$ will converge to the ground state of $H_{\text{eff}}$, and that can not be captured by a BTS in general. This is why the BTS wavefunction is not exact in this limit, even though its energy reaches the exact energy.


\bibliography{BTS}


\end{document}